\documentclass[journal]{IEEEtran}
\usepackage{amsmath,amsfonts}
\usepackage{algorithmic}
\usepackage{array}
\usepackage{textcomp}
\usepackage{stfloats}
\usepackage{url}
\usepackage{verbatim}
\usepackage{graphicx}
\usepackage{balance}
\usepackage[numbers,sort&compress]{natbib}
\usepackage[colorlinks,
            linkcolor=red,
            anchorcolor=blue,
            citecolor=green]{hyperref}
            
\usepackage{array}
\usepackage{multirow}
\usepackage{longtable}
\usepackage{rotating}
\usepackage{booktabs}
\usepackage{float}  
\usepackage{subfigure}

\usepackage{footnote}
 \usepackage{threeparttable}
 \usepackage{flushend}

\def\BibTeX{{\rm B\kern-.05em{\sc i\kern-.025em b}\kern-.08em
             T\kern-.1667em\lower.7ex\hbox{E}\kern-.125emX}}
\begin{document}
\title{ \textbf{BERT}: Accelerating Vital Signs Measurement for\\ \textbf{B}ioradar with An \textbf{E}fficient \textbf{R}ecursive \textbf{T}echnique }
  \author{Chengyao Tang, Yongpeng Dai, Zhi Li, Yongping Song, Fulai Liang, Tian Jin, \IEEEmembership{Member, IEEE}
				  \thanks{Manuscript created Nov. 1, 2023. This work was supported by the National Natural Science Foundation of China (Grant Nos. 61971430 and 62371454). (\textit{Corresponding author: Tian Jin}).}
				  \thanks{Chengyao Tang, Yongpeng Dai, Zhi Li, Yongping Song and Tian Jin are with the College of Electronic Science and Technology, National University of Defense Technology, Changsha 410073, China. \textit{(E-mail: \{cyt, dai\_yongpeng, lizhi, songyongping08, tianjin\}@nudt.edu.cn)}.
				  }%
                    \thanks{Fulai Liang is with the Shaanxi Key Laboratory for Bioelectromagnetic Detection and Intelligent Perception, School of Military Biomedical Engineering, Air Force Medical University, Xi'an 710032, China. \textit{(E-mail: liangfulai@fmmu.edu.cn)}.
				  }%
         }

  \markboth{Journal,~Vol.~18, No.~9, December~2023}%
  {How to Use the IEEEtran \LaTeX \ Templates} 

  \maketitle

\begin{abstract}
Recent years have witnessed the great advance of bioradar system in smart sensing of vital signs (VS) for human healthcare monitoring. As an important part of VS sensing process, VS measurement aims to capture the chest wall micromotion induced by the human respiratory and cardiac activities. Unfortunately, the existing VS measurement methods using bioradar have encountered bottlenecks in making a trade-off between time cost and measurement accuracy. To break this bottleneck, this letter proposes an efficient recursive technique (BERT) heuristically, based on the observation that the features of bioradar VS meet the conditions of Markov model. Extensive experimental results validate that BERT measurement yields lower time costs, competitive estimates of heart rate, breathing rate, and heart rate variability. Our BERT method is promising us a new and superior option to measure VS for bioradar. This work seeks not only to solve the current issue of how to accelerate VS measurement with an acceptable accuracy, but also to inspire creative new ideas that spur further advances in this promising field in the future.
\end{abstract}

\graphicspath{Figures}
\section{Introduction}\label{Introduction}
\begin{figure}[t]
	\centering
	\includegraphics[width=0.88\columnwidth]{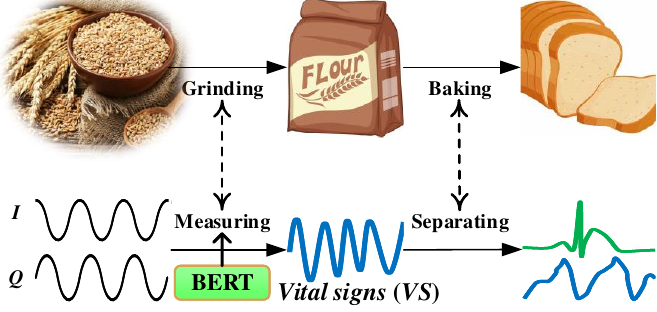}
	\caption{An figurative analogy for the task of bioradar-based VS measurement.\label{Fig1}}
\end{figure}

\IEEEPARstart{B}{ioradar} systems record both surface and internal motions of the human body by radiating and receiving electromagnetic waves \cite{zhang2019BioradarTechnologyRecent}, and enable many healthcare applications such as vital signs (VS) sensing, fall detection and cardiovascular dynamic monitoring \cite{paternianiRadarBasedMonitoringVital2023,schellenberger2020DatasetClinicallyRecorded,Amin2016RadarSignal,Mercuri2023BiomedicalRadar,Lauteslager2019CoherentUWB,Lauteslager2022DynamicMicrowave}. Since VS sensing including the estimation of breathing and heartbeat rates, is beneficial to various applications but is generally challenging due to the tiny amplitude of chest wall micromotion induced by the human cardiorespiratory activity, it is necessary to retrieve such micromotion containing VS from bioradar echoes, i.e., to achieve VS measurement \cite{park2007ArctangentDemodulationDC,wang2014NoncontactDistanceAmplitudeIndependent,xu2021LargeDisplacementMotion}. An efficient measurement can meet the long-term VS sensing requirement and provide high-quality raw materials for the subsequent separation of breathing and heartbeat, laying the groundwork for the estimation of breathing rate (BR), heart rate (HR) and heart rate variability (HRV). For example, as shown in Fig. (\ref{Fig1}), obtaining a good VS measurement result from the received in-phase (I) and quadrature-phase (Q) baseband signals can be likened to grinding wheat into fine-quality flour, while the subsequent procedures correspond to the process of baking bread. If we fail to acquire excellent flour, no matter how skilled the chef’s craftsmanship is in the subsequent processing, it would be challenging to create a delicious loaf of bread. Hence, measuring VS holds profound significance and captivates the researchers with its intriguing nature.
\par
It's known that the VS measurement principle of bioradar is to demodulate the phase values containing VS from the compensated IQ signals of target \cite{aardal2013PhysicalWorkingPrinciples,park2007ArctangentDemodulationDC,wang2014NoncontactDistanceAmplitudeIndependent,xu2021LargeDisplacementMotion}. The arctangent demodulation (AD) algorithm \cite{park2007ArctangentDemodulationDC} has the fastest processing speed, but it exhibits serious measurement errors attributed to the fact that it requires an unwrapping operation to solve the phase discontinuity and ambiguity in the measured VS result. To eliminate unwrapping operations in the AD algorithm and enable accurate VS measurement, Wang \emph{et al.} \cite{wang2014NoncontactDistanceAmplitudeIndependent} introduced differentiate and cross-multiply technique (DACM) based on the AD algorithm through a digital signal processing strategy which involved differentiation followed by integration. However, due to the integration requiring accumulators in the DACM algorithm, more computational resources is required, which makes it unsuitable for long-term VS measurement. Following the DACM algorithm, several modified versions \cite{xu2021LargeDisplacementMotion,zhang2022KalmanFilterCrossMultiply} have been provided, but they still essentially use the DACM algorithm to perform VS measurement task. From the perspective of the algorithm influence and feasibility, AD and DACM algorithms remain the mainstream solutions of VS measurement for bioradar currently, however, those two algorithms cannot make a trade-off between time cost and measurement accuracy, so it is necessary to explore a simple yet effective alternative.
\section{Motivation}\label{Motivation}
If we contemplate the bioradar-measured VS from the perspective of state correlation, a pivotal query emerges: Is there an efficient recursive technique available to characterize these measured VS values? This inquiry, both riveting and significant, takes center stage as the employment of recursive techniques emerges as a catalyst, greatly reducing the time cost of VS measurement task to seamlessly align with the demands of long-term VS sensing. The discernment of whether VS values can be elegantly articulated through recursive equations hinges upon an astute evaluation of their intrinsic characteristics. Actually, the VS amplitude is within a limited range (0.2mm $\sim$ 12mm) since the body surface micromotion is relatively small caused by cardiorespiratory activity \cite{paternianiRadarBasedMonitoringVital2023}, and this micromotion is rhythmic in normal conditions, implying that the current VS value could be predicted by the last one. A model in which all states are in a finite range and the current state depends on the previous states is mathematically known as Markov model \cite{MGM1}, which can establish a recursive relationship between the current state and the previous state. Obviously, the above features of bioradar VS meet the conditions of Markov model. If a recursive relationship could be constructed for the bioradar VS, the computational time would be greatly reduced. Unfortunately, there is no VS measurement method that provides this relationship for bioradar. Therefore, our motivation is to break this bottleneck by Markov model. To achieve fast VS measurement with an acceptable accuracy, we propose a heuristic method using bioradar with an efficient recursive technique (BERT), which employs the Markov-Gauss model to construct a state equation for the micromotion-induced phase information of bioradar echoes.
\section{Methods and rationale}
After performing a perfect IQ imbalance compensation on bioradar echo of the target, the output can be expressed as follows \cite{xu2021LargeDisplacementMotion}:
\begin{equation}
    \label{eq1}
    c_n=\left[ \begin{matrix}
	\cos \left( \theta +\frac{4\pi}{\lambda}x_n \right)&		\sin \left( \theta +\frac{4\pi}{\lambda}x_n \right)\\
\end{matrix} \right] ^{\mathrm{T}}, n \in \mathbb{N},
\end{equation}
where $\theta $ is the phase shift caused by the traveling distance of the electromagnetic waves. $x_n=x_n^h+x_n^b$ represents the cardiorespiratory micromotion to be measured, with the heartbeat signal $x_n^h$ and breathing signal $x_n^b$. $\lambda$ is the carrier wavelength of bioradar. By applying the rotation operation $\varGamma \left( \theta \right)=\left[ \begin{matrix}
	\cos \left( \theta  \right)&		\sin \left( \theta  \right)\\
	-\sin \left( \theta  \right)&		\cos \left( \theta  \right)\\
    \end{matrix} \right]$ with
\begin{equation}
    \label{eq2}
	\varGamma \left( \theta  \right) c_n
           =\left[ \begin{array}{c}
	\cos \left( \frac{4\pi}{\lambda}x_n \right)\\
	\sin \left( \frac{4\pi}{\lambda}x_n \right)\\
   \end{array} \right]
   \overset{\mathrm{def}}{=}\left[ \begin{array}{c}
	\cos \left( \varPhi _n \right)\\
	\sin \left( \varPhi _n \right)\\
\end{array} \right]
\overset{\mathrm{def}}{=}\left[ \begin{array}{c}
	I_n\\
	Q_n\\
\end{array} \right].\\
\end{equation}
\par
The next step is to measure VS contained in the micromotion-induced phase information $\varPhi _n$. As analyzed in the \hyperlink{Motivation}{\textit{Motivation}}, $\varPhi_n$ meets the conditions of the Markov model. Therefore, a Markov model with Gauss process \cite{MGM1,MGM2} is introduced to heuristically depict it as follows:
\begin{equation}
\label{eq3}
\varPhi _n=
\varPhi _m\text{e}^{-\beta \left( n-m \right)}
+G_m\sqrt{\frac{\sigma ^2}{2\beta}\left[ 1-\text{e}^{-2\beta \left( n-m \right) ^2} \right]},m<n
\end{equation} 
where $\beta$ is a discrete factor. A larger discrete factor means a low relevance between two neighbor values. $G$ expresses a Gauss process with zero mean and variance $\sigma ^2$.
\par
Considering that the sampling frequency of bioradar raw data is much higher than the vibration frequencies of human breathing and heartbeat activities ($<$ 1.7 Hz), the intervals and variations between neighboring sample values are very small in the bioradar VS, so that there exists high relevance between the current phase and the previous phase. Assuming that $\beta=0$ and $m=n-1$, the Markov-Gauss model (MGM) can be simplified as follows:
\begin{equation}
    \label{eq4}
    \varPhi _n=\varPhi _{n-1}+\sigma G_{n-1}=\varPhi _{n-1}+\mathcal{G}_{n-1}.
\end{equation}
The expectation and variance of MGM are given by
\begin{equation}
    \label{eq5}
    \begin{cases}
	  \mathrm{E} \left( \varPhi _n \right) 
       =\mathrm{E} \left( \varPhi _{n-1}+\mathcal{G}_{n-1} \right) 
       =\mathrm{E} \left( \varPhi _{n-1} \right)\\
	\mathrm{D} \left( \varPhi _n \right) 
       =\mathrm{D} \left( \varPhi _{n-1}+\mathcal{G}_{n-1} \right) 
       =\mathrm{D} \left( \varPhi _{n-1} \right) +\sigma ^4\\
\end{cases}.
\end{equation}
\par
To formulate the relationship of phases precisely, obtaining an exact model of the Gauss process is essential. To this end, we model $\mathcal{G}_{n-1}$ as follows:
\begin{equation}
    \label{eq6}
    \begin{aligned}
		\mathcal{G} _{n-1}
        &=\varPhi _n-\varPhi _{n-1}\\
		&\approx \sin \left( \varPhi _n-\varPhi _{n-1} \right) \\
		&=\sin \left( \varPhi _n \right) \cos \left( \varPhi _{n-1} \right) 
		-\cos \left( \varPhi _n \right) \sin \left( \varPhi _{n-1} \right)  \\
		&=\left( Q_nI_{n-1}-I_nQ_{n-1} \right) /\left( I_{n}^{2}+Q_{n}^{2} \right) \\
        &\overset{\mathrm{def}}{=}M_{n-1}.\\
    \end{aligned}
\end{equation}
The expectation and variance of the new MGM are below:
\begin{equation}
    \label{eq7}
    \begin{cases}
	  \mathrm{E} \left( \varPhi _{n-1}+M_{n-1} \right) =\mathrm{E} \left( \varPhi _{n-1} \right)\\
	\mathrm{D} \left( \varPhi _{n-1}+M_{n-1} \right) 
       =\mathrm{D} \left( \varPhi _{n-1} \right) +\varUpsilon\\
    \end{cases},
\end{equation}
where $\varUpsilon =0.5+2\sin \left( \varPhi _n-\varPhi _{n-1} \right) \mathrm{E} \left( \varPhi _{n-1} \right)$ and is not a constant as follows:
\begin{equation}
    \label{eq8}
    \varUpsilon \in 
    \left[ 0.5-2\mathrm{E} \left( \varPhi _{n-1} \right) ,
    0.5+2\mathrm{E} \left( \varPhi _{n-1} \right) \right].
\end{equation}
Eq. (\ref{eq8}) indicates that the variance of new MGM is in a moving range, so this model can tolerate abrupt phase variations.
\par
Using the new MGM, a recursive expression (BERT) of the bioradar echo phases carrying VS is obtained as follows:
\begin{equation}
    \label{eq9}
       \varPhi _n=\varPhi _{n-1}+M _{n-1},\;\;
       \varPhi _0=\arctan\left(Q_0/I_0\right).
\end{equation}
\par
\begin{figure}[t]
\vspace{-0.30cm}
	\centering
	\includegraphics[width=0.88\columnwidth]{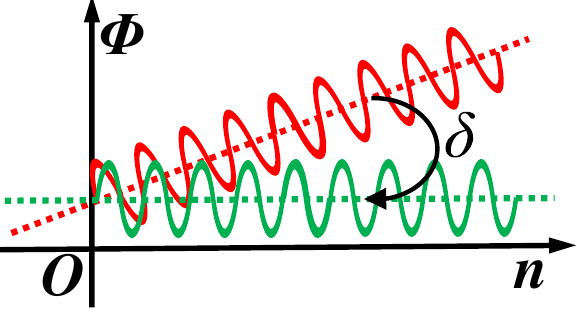}
	\caption{Angular transform to compensate the linear error caused by recursive accumulation.\label{Fig2}}
 \vspace{-0.20cm}
\end{figure}

Since the recursive technique is prone to linear drift as shown in Fig. \ref{Fig2}, an angular transform ($\varPhi_n \cos \delta $) is used to compensate for this linear error and make the measurement more accurate.
\section{Experiments}
\begin{figure}[h]
	\centering
	\includegraphics[width=0.88\columnwidth]{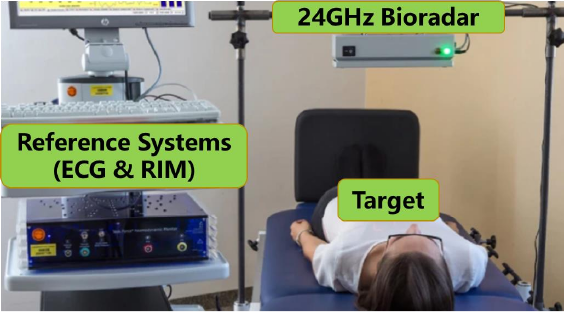}
	\caption{Experimental setup \cite{schellenberger2020DatasetClinicallyRecorded}. The bioradar, suspended on a stand, emits signals towards a target lying motionless on a bed and receives echoes to measure the VS.\label{Fig3}}
\end{figure}
The bioradar dataset with synchronized reference signals \cite{schellenberger2020DatasetClinicallyRecorded} from Germany is applied to validate VS measurement performance, and its experimental setup is shown in Fig. \ref{Fig3}. The bioradar system used in this dataset is a 24 GHz continuous-wave (CW) radar based on the advanced Six-Port interferometry technology. The reference system comprises a set of electrocardiogram (ECG) equipment and a respiratory impedance monitor (RIM), which can keep tight synchronization with the bioradar system. More details of this dataset can be found in Ref. \cite{schellenberger2020DatasetClinicallyRecorded}.
\begin{figure}[h]
\vspace{-0.30cm}
	\centering
	\includegraphics[width=0.95\columnwidth]{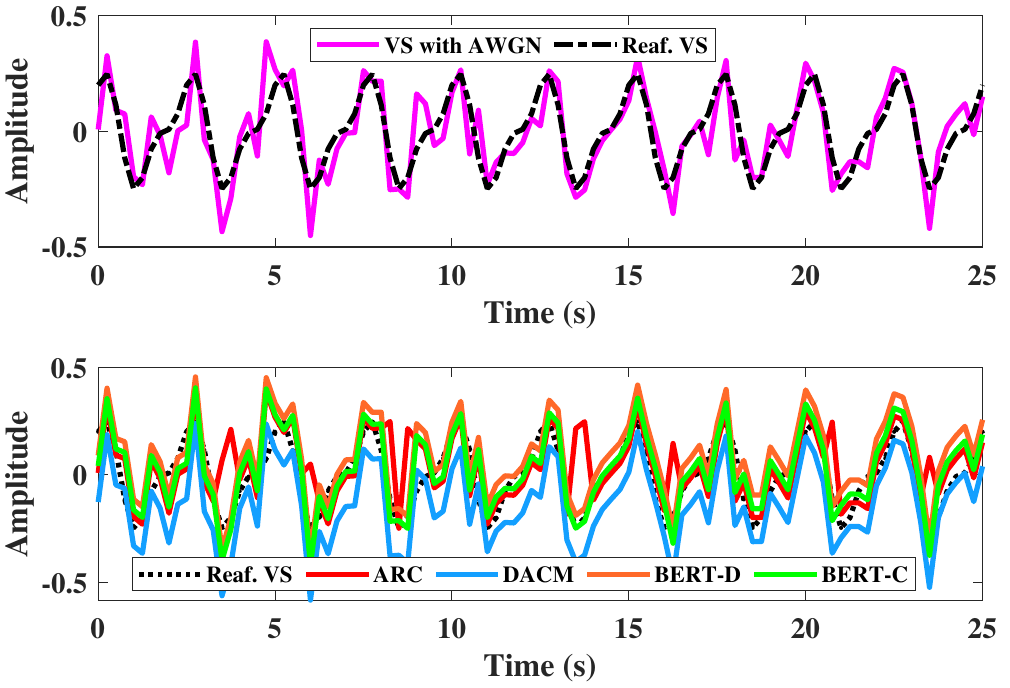}
	\caption{Simulation results of various measurement methods in VS amplitudes (VSAs). BERT-D = BERT with draft; BERT-C = BERT without draft by compensation. \label{Fig4}}
 \vspace{-0.30cm}
\end{figure}
\begin{figure}[h]
\vspace{-0.30cm}
	\centering
	\includegraphics[width=0.95\columnwidth]{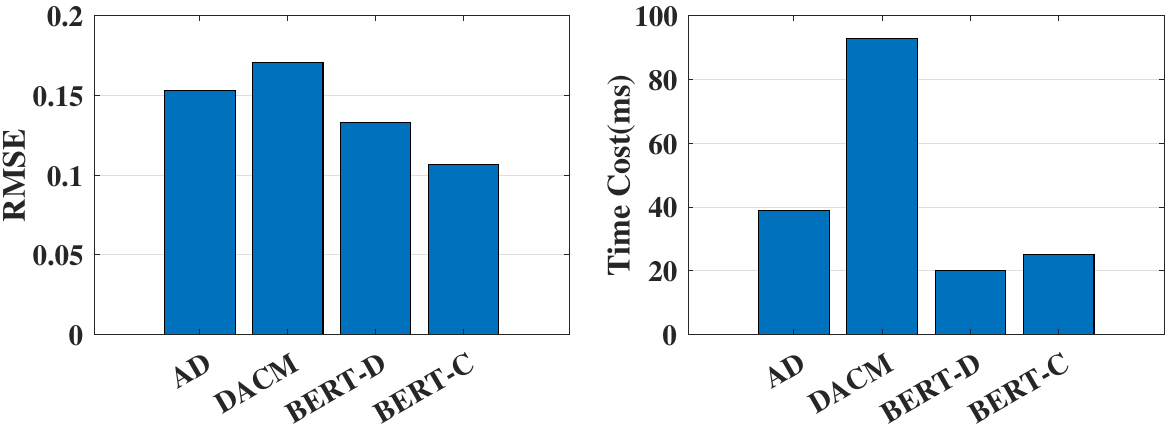}
	\caption{Simulation results of various measurement methods contain time costs and root mean square errors (RMSEs), respectively.\label{bar2}}
 \vspace{-0.30cm}
\end{figure}
\par
First, simulation results are displayed in Figs. \ref{Fig4} and \ref{bar2}, which show that the BERT measurement provides faster and more accurate performance than other methods. 
\par
\begin{figure}[t]
\vspace{-0.30cm}
	\centering
	\includegraphics[width=0.88\columnwidth]{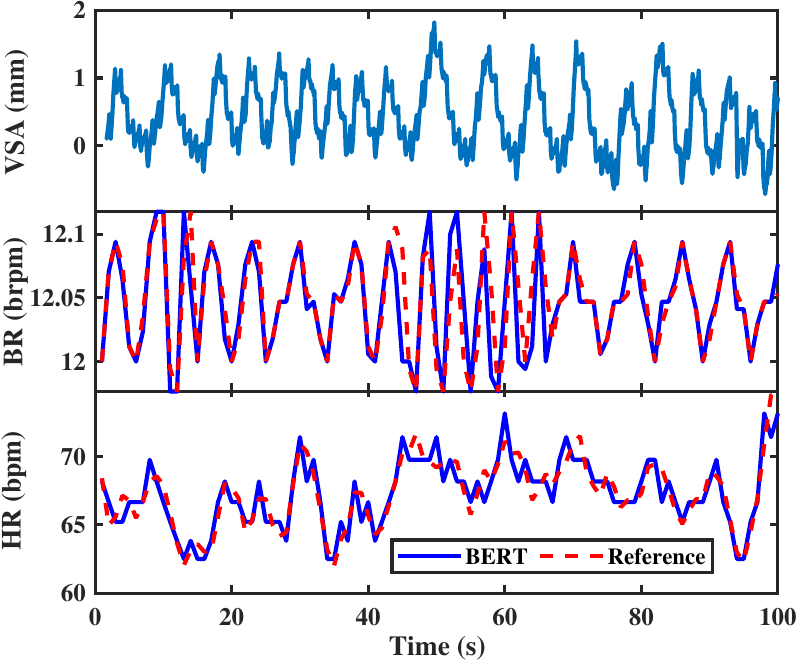}
	\caption{Real results of BERT measurement.\label{Fig5}}
 \vspace{-0.30cm}
\end{figure}
Second, a target (T1) is randomly selected from the bioradar dataset \cite{schellenberger2020DatasetClinicallyRecorded} to testify the real performance of BERT measurement, and the results is shown in Fig. \ref{Fig5}. It can be seen that bioradar can effectively acquire the BR and HR values that are remarkably close to those of the reference systems (RIM and ECG). 
\par
\begin{figure}[h]
\vspace{-0.30cm}
	\centering
	\includegraphics[width=0.88\columnwidth]{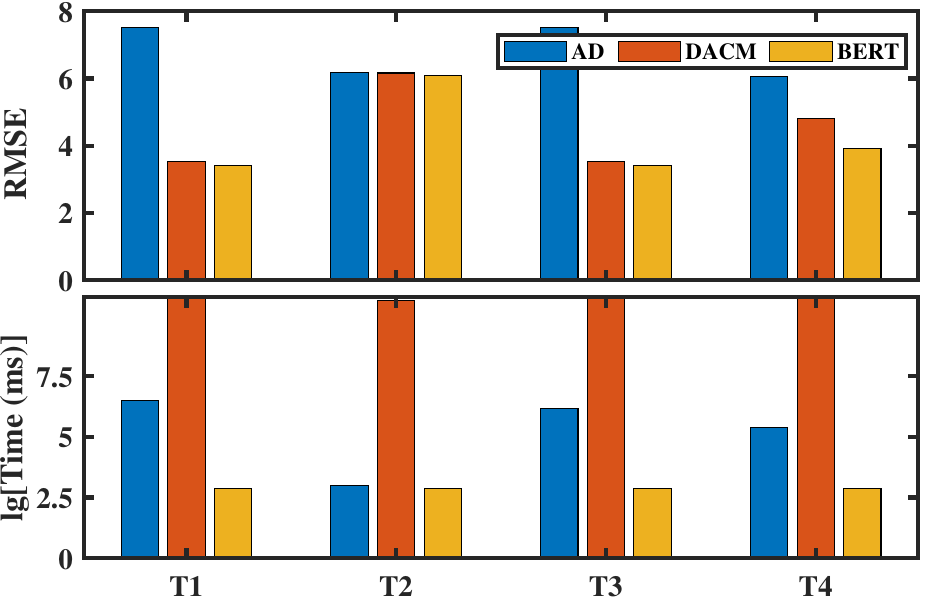}
	\caption{HR RMSEs and time costs measured with various methods for 4 targets.  The BR RMSEs are less than $10^{-4}$ and hence not shown here.\label{Fig7}}
 \vspace{-0.30cm}
\end{figure}
Third, in order to offer more evidence, supplementary experiments are conducted with randomly selecting three additional targets (T2$\sim$T4). Fig. \ref{Fig7} verifies that BERT method can achieve a 19-fold and 46-fold speedup compared to the AD and DACM methods, respectively, while yielding the smallest measurement error. 
\par
\begin{figure*}[t]
\vspace{-0.30cm}
\centering
\includegraphics[width=0.98\textwidth]{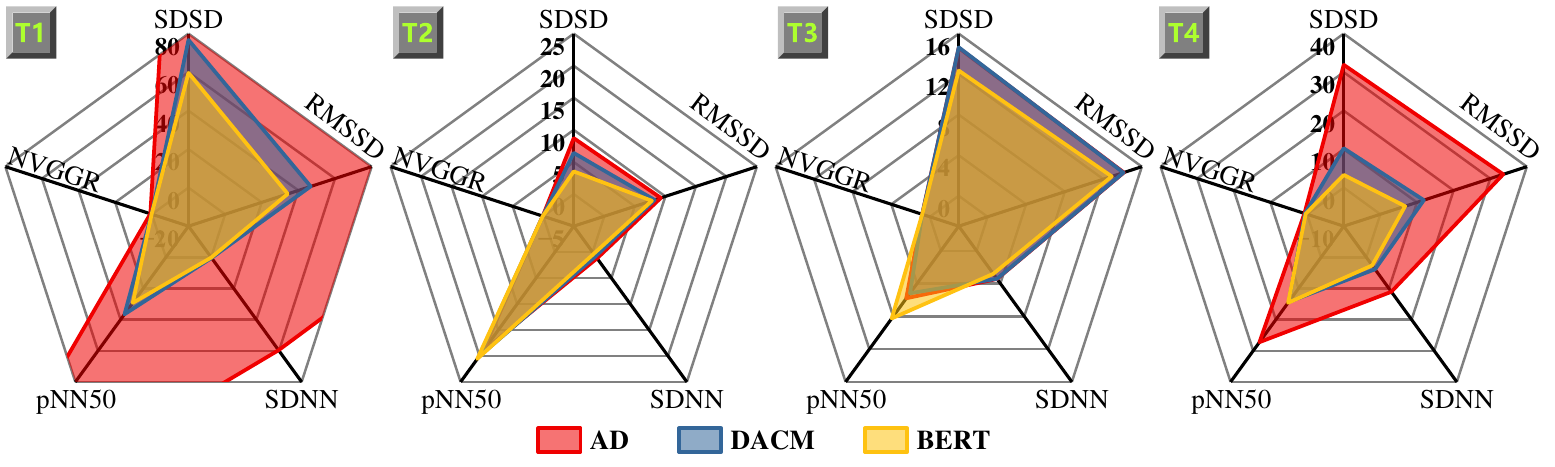}
\caption{Radial plots of relative errors of HRV indexes (NNVGR $=$ Mean RR intervals, SDNN $=$ Standard deviation of RR intervals, RMSSD $=$ Root mean square of differences between adjacent RR intervals; SDSD $=$ Standard deviation of differences between adjacent RR intervals, pNN50 $=$ The percentage of successive RR intervals that differ more than 50ms). Smaller shadow areas in the radial plot mean better HRV measurement performance.\label{Fig8}}
\vspace{-0.30cm}
\end{figure*}

Finally, compared to the accuracy of HRV estimates obtained by using various VS measurement techniques, BERT method presents the smallest shadow area in the radial plots of relative errors of HRV indexes for all targets, as shown in Fig. \ref{Fig8}.
\section{Conclusions}
To break the dilemma of mutual restriction between measurement accuracy and time cost for bioradar-based VS measurement task, BERT method is proposed in this work. It utilizes the Markov-Gauss model to construct a simple yet effective state equation, empowering the precise and fast measurement of vital signs. Extensive experimental results validate that BERT measurement yields low time costs, competitive estimates of BR, and HR and HRV. Hence, our BERT method holds promise as a novel and enhanced choice for VS measurement through bioradar technology, and future iterations have the potential to showcase even more substantial performance advantages.
\par
\textbf{Limitation and Future Work.} Our method is heuristic and uses a coarse-grained scheme to compensate the recursive error of the BERT measurement. In the future, we will further explore model-calibration approaches that can be jointly optimized with the BERT method to achieve finer-grained VS measurement for bioradar systems.
  \bibliographystyle{IEEEtran}


\end{document}